\documentclass[nofootinbib, reprint, preprintnumbers, amsmath, amssymb, amsfonts, aps, pra, superscriptaddress, floatfix]{revtex4-2}
\usepackage[utf8]{inputenc}
\usepackage{mathtools}
\usepackage{graphicx}
\usepackage{dcolumn}
\usepackage{bm}
\usepackage{physics}
\usepackage{color}
\usepackage{braket}
\usepackage[normalem]{ulem}
\usepackage{enumerate}
\usepackage{enumitem}
\usepackage[colorlinks=true,linkcolor=blue,citecolor=blue,urlcolor =blue]{hyperref}
\usepackage{mathrsfs}
\usepackage{bbm}
\usepackage{siunitx}

\usepackage{comment}
\usepackage{lipsum}

\begin{document}
\title{Proposal to experimentally evaluate the adiabatic condition of quantum annealing in coupled systems of Kerr parametric oscillators}

\author{Yuichiro Mori}
\thanks{These authors equally contributed to this paper.}
\email{mori-yuichiro.9302@aist.go.jp}
\affiliation{Global Research and Development Center for Business by Quantum-AI Technology (G-QuAT), National Institute of Advanced Industrial Science and Technology (AIST), 1-1-1, Umezono, Tsukuba, Ibaraki 305-8568, Japan}%

\author{Harunobu Hiratsuka}
\thanks{These authors equally contributed to this paper.}
\affiliation{Department of Electrical, Electronic, and Communication Engineering, Faculty of Science and Engineering, Chuo university, 1-13-27, Kasuga, Bunkyo-ku, Tokyo 112-8551, Japan}%

\author{Yuichiro Matsuzaki}
\email{ymatsuzaki872@g.chuo-u.ac.jp}
\affiliation{Department of Electrical, Electronic, and Communication Engineering, Faculty of Science and Engineering, Chuo university, 1-13-27, Kasuga, Bunkyo-ku, Tokyo 112-8551, Japan}%

\date{\today}


\begin{abstract}
Quantum annealing (QA) is 
an algorithm to find the ground state of the problem Hamiltonian by using an adiabatic time evolution. An approach to 
evaluate adiabaticity in the experiment by applying spectroscopic techniques has recently been 
suggested. However, this 
method requires temporal oscillation of interaction strength during 
QA, posing challenges for experimental demonstration. Here, we propose an experimental method for 
evaluating adiabaticity when performing 
QA with a parametric oscillator with Kerr nonlinearity (KPO). Importantly, our proposal offers a significant advantage by eliminating the need for temporal oscillation of interactions during 
QA. We 
investigate its performance through numerical simulations, and we show the feasibility of our method. 
\end{abstract}

\maketitle
\section{Introduction}
Quantum annealing (QA)~\cite{Kadowaki_1998_pre,farhi2000quantum}, a method for solving combinatorial optimization problems based on quantum 
mechanics, has garnered significant attention. In QA, 
we prepare the ground state of a 
simple Hamiltonian, which is called the driver Hamiltonian, and gradually 
change the Hamiltonian into a problem Hamiltonian 
whose ground state corresponds to the solution. This approach has been explored from various 
point of view, including computational speed~\cite{Somma_2012PRL,Muthukrishnan_2016PRX,Hastings_2021QUANTUM}, implementation techniques~\cite{Imoto_seki_2022,Miyazaki_2022PRA}, and algorithms ~\cite{Roland_2002PRA,Chang_2022PRXQ,Schiffer_2022_PRXQ}. Moreover, 
D-wave developed a QA machines with thousands of superconducting qubits, and there are many demonstrations in fields such as quantum chemistry~\cite{Babbush_2014_scirep,Teplukhin_2020SciRep}, machine learning \cite{Benedetti_2017_prx,Prasanna_2021SciRep}, and high-energy physics~\cite{Mott_2017_nature}.

The core principle that enables 
QA to solve problems is the ``adiabatic theorem'' in quantum mechanics. According to this theorem, the system remains in a ground state as long as the dynamics is adiabatic. 
Consequently, maintaining the adiabatic condition is crucial to obtain the correct answer by using QA. 

The adiabaticity of the 
dynamics is 
assessed using the following quantitative conditions~\cite{Morita_2008JMP, mori2022evaluate,Albash_2018}:
\begin{align}
     \frac{|\braket{m(s)|\dot{\mathcal{H}}(s)|0(s)}|}{|E_{m}(s)-E_{0}(s)|^{2}}\ll T_{\rm ann}, \label{eq:adiabatic_criterion}
\end{align}
where $T_{\text{ann}}$ is the annealing time, $s=t/T_{\text{ann}}$ is a dimensionless time normalized by the annealing time, 
$|m(s)\rangle$ ($|0(s)\rangle$) represents the $m$-th excited (ground) state. 
$\dot{H}(s)$ is the derivative of the instantaneous Hamiltonian at time $s$, and $E_m(s)$ ($E_0(s)$) is the energy of the $m$-th excited state (ground state).
This is called the adiabatic condition, and is derived by considering the first-order term in the perturbation expansion where 
higher-order terms are ignored~\cite{Morita_2008JMP}.
Recently, a more rigorous analysis of the adiabaticity for QA was done
by considering the higher order~\cite{kimura2022rigorous}.

For practical applications, it is not straightforward to check whether the adiabatic condition is satisfied during practical QA.
Theoretical evaluation of the adiabatic condition requires diagonalizing the Hamiltonian. However, if this were possible, it would imply that the original combinatorial optimization problem could be solved directly, negating the need for 
QA. Therefore, for the applications in real-world problems, it is preferable to assess such adiabaticity experimentally during the implementation of QA. 
Recently, a method has been proposed to achieve this~\cite{mori2022evaluate}. In this 
method, Rabi oscillations are induced between the ground and excited states using an oscillatory field during 
QA. 
The power spectrum is then obtained by performing the Fourier transform of the signal in the time domain of the Rabi oscillation.
Finally, the information from the obtained data is processed by a classical computer to estimate the values of the numerator and denominator of the adiabatic condition in Eq~\eqref{eq:adiabatic_criterion}. However, 
this methods requires an oscillation of the coupling strength, making experimental demonstration challenging.

On the other hand, 
QA using Kerr 
parametric oscillators (KPOs) has recently been proposed~\cite{goto2016bifurcation,puri2017engineering}. 
KPOs 
is a 
bosonic system with large Kerr nonlinearity and a parametric drive. When the parametric drive is strong, a bifurcation process occurs, and we can use this system as a qubit~\cite{milburn1991quantum,wielinga1993quantum}, allowing for gated quantum computation and QA~\cite{milburn1991quantum,wielinga1993quantum, Cochrane1999_PRA,meaney2014quantum,goto2016bifurcation,puri2017engineering,yamaji2023correlated}. 
We can realize the KPO 
by introducing Josephson junctions in superconducting resonators~\cite{meaney2014quantum}.
 KPO-based quantum bits have already been realized experimentally~\cite{Grimm2020_nature}. 
It is worth mentioning that QA with KPOs does not require to change the interaction strength between KPOs~\cite{goto2016bifurcation}, while we change the interaction strength between qubits in the conventional QA~\cite{kadowaki1998quantum}.

Here, we propose a method to experimentally 
evaluate the adiabatic condition of QA with KPOs.
Unlike 
previous methods~\cite{mori2022evaluate}, 
our approach has the significant advantage of not requiring temporal oscillation of the interaction strength during QA. 
This comes from the fact that the interaction strength can be fixed during QA with KPOs.
We have evaluated its performance through numerical simulations and will discuss the results.
\section{ quantum annealing}
We introduce 
QA~\cite{Kadowaki_1998_pre,farhi2000quantum}. QA employs the following Hamiltonian:

\begin{align}
    \mathcal{H}_{\mathrm{conv}}(s) =f(s) \mathcal{H}_{\mathrm{D}} + (1-f(s)) \mathcal{H}_{\mathrm{P}},\label{def:QA_Hamiltonian_conv}
\end{align}
where $\mathcal{H}_{\mathrm{D}}$ denotes the driver Hamiltonian, 
$\mathcal{H}_{\mathrm{P}}$ denotes the problem Hamiltonian, 
$s=t/T_{\rm{ann}}$ denotes the normalized 
time, $T_{\rm{ann}}$ denotes the annealing time, and 
$f (s)$ denotes 
a function satisfying $f(0) = 1$ and $f(1) = 0$.
Throughout the paper, $f(s)$ is considered as follows.
\begin{align}
    f(s)&=1-s \label{eq:ann_linsche}
\end{align}
The Hamiltonian at $s=0$ ($s=1$) is the driver (problem) Hamiltonian $\mathcal{H}_{\mathrm{D}}$ ($\mathcal{H}_{\mathrm{P}}$). 
After preparing the ground state of the driver Hamiltonian at $s=0$, we change the Hamiltonian based on the scheduling as mentioned above. 
According to the adiabatic theorem, if the annealing time \( T_{\text{ann}} \) is sufficiently long, 
the system remains in the ground state.
Typically, the driver Hamiltonian and problem Hamiltonian are chosen as follows:
\begin{align}
&    \mathcal{H}_{\mathrm{D}}=\sum _j B \hat{\sigma}^{(j)}_x 
\label{convqanoninteract}
\\
&    \mathcal{H}_{\mathrm{P}}=\sum _j h_j \hat{\sigma}^{(j)}_z+ \sum _{i,j}J_{i,j}
\hat{\sigma}^{(i)}_z\hat{\sigma}^{(j)}_z
\label{convqainteract}
\end{align}
where $\hat{\sigma}^{(j)}_z$ ($\hat{\sigma}^{(j)}_x$) is the Pauli matrix acting on the $j$-th qubit for $z$ ($x$)
direction, $B$ is the amplitude of the uniform transverse magnetic field, $h_j$ is the horizontal magnetic field for the $j$-th qubit, $J_{ij}$ is the coupling strength between the $i$-th qubit and $j$-th qubit.
\section{ Kerr Nonlinear Parametric Oscillator}
In this section, we describe Kerr Parametric Oscillators (KPOs). The KPOs are bosonic systems featuring the bifurcation phenomena induced by the parametric drive and Kerr non-linearity, 
which has recently garnered significant attention as devices for quantum computing~\cite{milburn1991quantum,wielinga1993quantum} 
and 
QA~\cite{goto2016bifurcation,puri2017engineering}.
Initially, we discuss single KPO and subsequently introduce a network of KPOs. 
In a coordinate system rotating at half the parametrically driven pump frequency 
with the rotating wave approximation, the Hamiltonian of a single KPO can be expressed as:
\begin{align}
    \mathcal{H}&=\chi \hat{a}^{\dagger 2}\hat{a}^{2} + \Delta \hat{a}^{\dagger} \hat{a}
    - p(\hat{a}^{2} + \hat{a}^{\dagger 2}) + r(\hat{a}+\hat{a}^{\dagger})\label{def:KPOHamil}
\end{align}
where \( \chi \), \( \Delta \), \( p \), and \( r \) represent the Kerr nonlinearity, detuning, parametric drive 
amplitude, and coherent drive 
amplitude, respectively.
If we set $p=r=0$, we obtain
\begin{align}
\mathcal{H}&=\chi \hat{a}^{\dagger 2}\hat{a}^{2} + \Delta \hat{a}^{\dagger} \hat{a}\label{def:KPOHamil_przero}.
\end{align}
Let us explain how to generate the so-called cat state by using an adiabatic operation. By setting $\Delta=r=0$, we obtain
\begin{align}
\mathcal{H} =\chi\left(\hat{a}^{\dagger 2}-\frac{p}{\chi}\right)\left(\hat{a}^{2}-\frac{p}{\chi}\right)-\frac{p^{2}}{\chi}, \label{eq:nodetuning}
\end{align}
and the ground state resides in the eigenspace spanned by the two coherent states \( | \frac{p}{\chi} \rangle \) and \( | -\frac{p}{\chi} \rangle \). 
A coherent state is defined as
\begin{align}
\ket{\alpha} = e^{-\frac{|\alpha|^{2}}{2}}\sum_{k} \frac{\alpha^{k}}{\sqrt{k!}}\ket{k} \label{eq:coherent_state}
\end{align}
where \( |k\rangle \) is the Fock state. 
Here,
\( |\sqrt{p/\chi\rangle} \) and \( |-\sqrt{p/\chi\rangle} \) are orthogonal in the limit of large \( p \), and so these two states can be used as qubits~\cite{cochrane1999macroscopically}.
We can prepare a vacuum state with the Hamiltonian of Eq.~\eqref{def:KPOHamil_przero}. By adiabatically changing the Hamiltonian from Eq.~\eqref{def:KPOHamil_przero} to Eq.~\eqref{eq:nodetuning}, it is possible to generate the cat state described as $\frac{1}{\sqrt{2}}(| \sqrt{\frac{p}{\chi}} \rangle +|-\sqrt{\frac{p}{\chi}} \rangle)$~\cite{milburn1991quantum,wielinga1993quantum}.

Next, we describe QA using KPO networks. For 
QA with multiple KPO Hamiltonians, also known as KPO networks, the following problem and driver Hamiltonians are employed.
\begin{align}
\mathcal{H}_\mathcal{P} = &\sum_{j=1}^{K} \left(\chi_{j}\hat{a}^{\dagger 2}_{j}\hat{a}_{j}^{2} + \Delta_{j}\hat{a}^{\dagger}_{j}\hat{a}_{j}-p_{j}(\hat{a}_{j}^{2}+\hat{a}^{\dagger 2}_{j})+r_{j}(\hat{a}_{j} + \hat{a}^{\dagger}_{j})\right)\nonumber\\
&\ +\sum_{j>j'}^{K} \left(J_{jj'}\hat{a}^{\dagger}_{j}\hat{a}_{j'}+J_{jj'}^{\ast}\hat{a}^{\dagger}_{j'}\hat{a}_{j}\right)\label{eq:KPO_Hp}\\
\mathcal{H}_\mathcal{D} = &\sum_{j=1}^{K} \left(\chi_{j}\hat{a}^{\dagger 2}_{j}\hat{a}_{j}^{2} + \Delta_{j}\hat{a}^{\dagger}_{j}\hat{a}_{j}+r_{j}(\hat{a}_{j} + \hat{a}^{\dagger}_{j})\right)\nonumber\\
&\ +\sum_{j>j'}^{K} \left(J_{jj'}\hat{a}^{\dagger}_{j}\hat{a}_{j'}+J_{jj'}^{\ast}\hat{a}^{\dagger}_{j'}\hat{a}_{j}\right)\label{eq:KPO_Hd}
\end{align}
where \( K \) denotes the number of KPOs, \( J_{jj'} \) denotes the coupling strength between KPOs, $\chi_j$ denotes the Kerr nonlinearity, $\Delta_j$ denotes the detuning, $p_j$ denotes the parametric drive amplitude, and $r_j$ denotes the coherent drive amplitude.

In this paper, we assume that 
we fix the values of \( \chi_j \) and \( J_{jj'} \) while we can change the values of \(\Delta_{j}\), \(p_j\), and \(r_j\) 
during the experiment. 

By setting appropriate detuning without coherent drive, the ground state of the driver Hamiltonian can be a vacuum state~\cite{goto2016bifurcation},
and we use this as the initial state for QA.
We adiabatically change the Hamiltonian from $\mathcal{H}_\mathcal{D}$ to $\mathcal{H}_\mathcal{P}$, and obtain the ground state of $\mathcal{H}_\mathcal{P}$. 
When we use \( |\sqrt{p/\chi}\rangle \) and \( |-\sqrt{p/\chi}\rangle \) as qubits for large $p$, this problem Hamiltonian is known to correspond to the Ising Hamiltonian~\cite{goto2016bifurcation}. Therefore, the ground state of the Ising Hamiltonian can be obtained~\cite{goto2016bifurcation,puri2017engineering}..



\section{Experimental Methods for Determining Adiabatic Conditions}
We introduce an experimental method to 
judge whether the dynamics of a given quantum system is adiabatic~\cite{mori2022evaluate}. More specifically, this method allows us to experimentally 
evaluate Eq.~\eqref{eq:adiabatic_criterion}, where the denominator represents the energy gap of the Hamiltonian \(\mathcal{H}(s) \) and the numerator is the transition matrix element of \( \mathcal{\dot{H}}(s) \). The following procedure allows us to measure these quantities from experiments. First, we adopt the following Hamiltonian:

\begin{align}
    \mathcal{H}(s) &=\mathcal{H}_{\rm{QA}}(s)
    +\mathcal{H}_{\mathrm{ext}}(s)\label{def:QA_Hamiltonian}
    \\
\mathcal{H}_{\rm{QA}}(s)&= A(s) \mathcal{H}_{\mathrm{D}} + (1-A(s)) \mathcal{H}_{\mathrm{P}}
    \\
    \mathcal{H}_{\mathrm{ext}}(s) &= \lambda (s) \dot{\mathcal{H}}_{\rm{conv}}(s_{1})\cos\left(\omega T_{\rm ann}  (s-s_1)\right)\label{def:external_Hamiltonian}
\end{align}
where \( \mathcal{H}_D \) (\(\mathcal{H}_P \)) is the driver (problem) Hamiltonian 
for the QA Hamiltonian of \( \mathcal{H}_{\text{QA}}(s) \), \( A(s) \) is the scheduling function, \( \mathcal{H}_{\text{ext}}(s) \) is the external drive Hamiltonian, \( \omega \) is the angular frequency of the drive field, \( \lambda(s) \) is the drive field amplitude, and \( s_1 \) is the dimensionless time to start the external drive.
\begin{figure}[h!]
    \centering
    \includegraphics[width = 8.5cm]{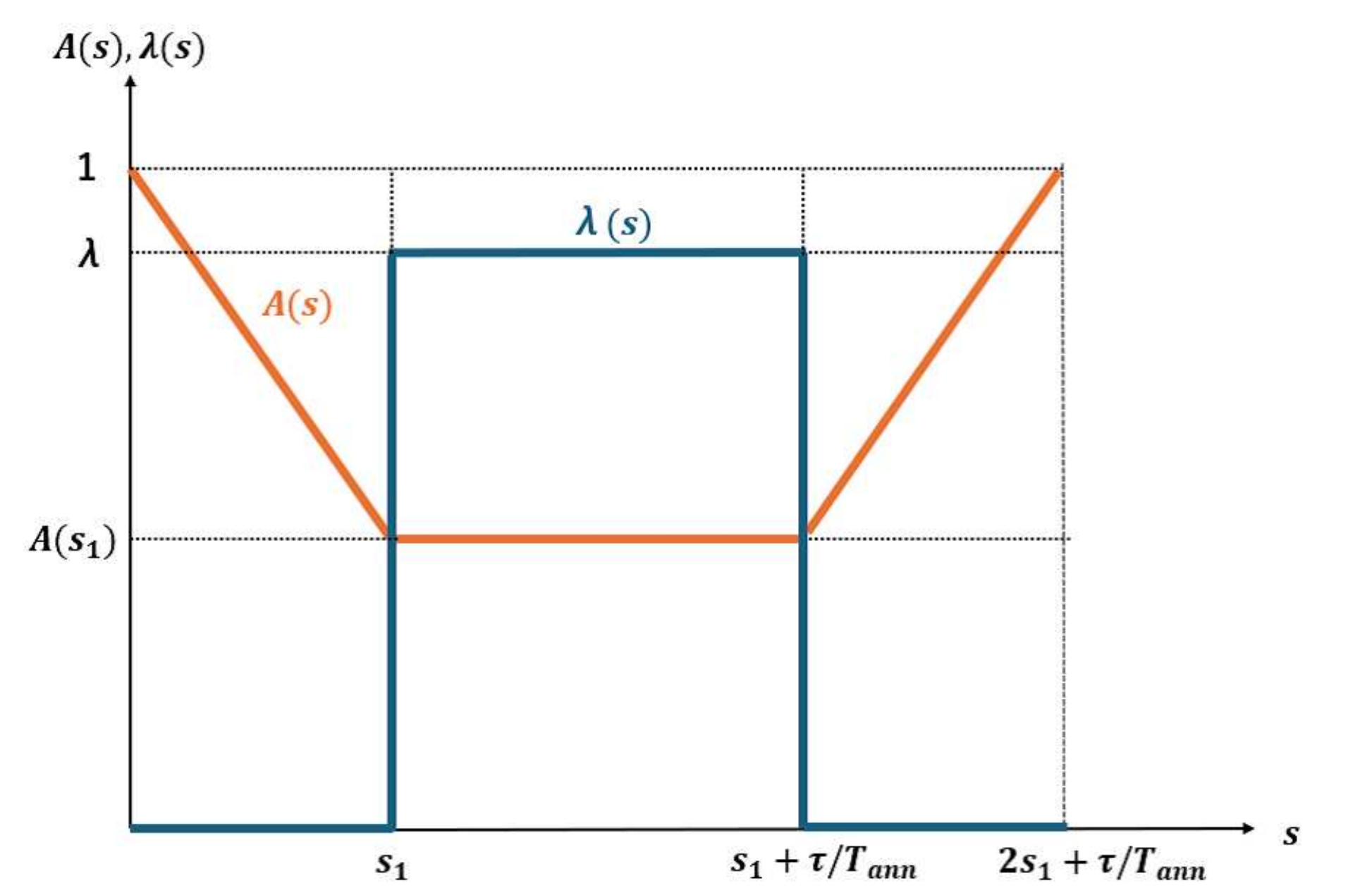}
\caption{
 Scheduling of the 
 parameters in the previous method to evaluate the adiabatic condition. 
}\label{zu_posuta.1}
\end{figure}
In Figure 1, \( A(s) \) and \(\lambda(s) \) are plotted against time, where \(A(s) = f(s) \) for $0\leq s\leq s_{1}$.

(1) First, prepare the ground state of the driver Hamiltonian $\ket{0(s=0)}$.

(2) Then the Hamiltonian $\mathcal{H}_{\rm QA}(s)$ is varied from $s=0$ to $s=s_{1}$ while setting $\lambda(s) = 0$, allowing the system to evolve in time by the QA Hamiltonian.

(3) 
Add an 
external driving
term to the Hamiltonian by setting \( \lambda(s) = \lambda \) at \( s = s_1 \). Let \( A(s) = f(s_1) \) and let the system evolve in time in this Hamiltonian for $s_{1}<s\leq s_{1}+\tau/T_{\rm ann}$.

(4) 
Turning off the drive as \( \lambda(s) = 0 \) at \( s = s_1 + \tau/T_{\text{ann}} \), and gradually change the Hamiltonian from \( H_{\text{QA}}(s_1) \) to \( H_D \) during \( s_1 + \tau/T_{\text{ann}} < s \leq 2s_1 + \tau/T_{\text{ann}} \). 
This corresponds to an operation called 
reverse 
QA~\cite{ohkuwa2018reverse,yamashiro2019dynamics}.

(5) We perform a projection measurement on the \( m \)-th excited state \( \ket{m(s = 0)} \) of the driver Hamiltonian and record the measurement results.

(6) Finally, these steps are repeated many times with different \( \omega \), \( s_1 \) and \( \tau \) . The probability obtained from these steps is defined as $p_{0,m}(\omega, s_1, \tau)$. In an actual experiment, we describe how to realize the third step with $\mathcal{H}_{\rm ext}(s)$.
The external driving Hamiltonian is written as
\begin{align}
\mathcal{H}_{\mathrm{ext}}(s)= \lambda
\dot{f}(s_1)\cos\left(\omega T_{\rm ann}(s-s_1)\right)\left( \mathcal{H}_{\mathrm{D}}
-\mathcal{H}_{\mathrm{P}}\right).
\label{oscillate}
\end{align}
Substituting 
Eqs.~\eqref{convqanoninteract} and \eqref{convqainteract} into 
Eq.~\eqref{oscillate}, we see that transverse magnetic fields, horizontal magnetic fields, and coupling strength should be temporally modulated. However, it is worth mentioning that such a temporal modulation of the coupling strength is difficult to realize for most of the systems. 

Let us explain why we can evaluate the adiabatic condition by using the procedure described above. 
For simplicity, we assume that the dynamics in the second and fourth steps are adiabatic. 
For a more general case, please refer to 
Ref.~\cite{mori2022evaluate}. 
for the sake of simple notation, we omit the expression ``$(s_1)$''. In our proposal, the measurement is done by sweeping time \( \tau \). Therefore, in this section, unless otherwise noted, we treat \( \tau \) as a variable.


We can diagonalize the QA Hamiltonian as follows: 
\begin{align}
    \mathcal{H}_{\mathrm{QA}} = \sum_{i}E_{i}\ket{i}\bra{i},\label{eq:diag_qa_hamito}
\end{align}
where \( E_i \leq E_j \) is satisfied for \( i < j \). 
Throughout the paper, we set 
\( \hbar = 1 \). 
Then, by moving to the rotating frame, 
the state of the system can be written as
\begin{align}
    \ket{\tilde{\psi}(\tau)}=e^{i\tilde{r}\tau\mathcal{H}_{\mathrm{QA}}}\ket{\psi(\tau)},\label{eq:transformation}
\end{align}
where $\tilde{r}$ denotes a proportional constant of the frequency of the rotating frame. The Hamiltonian in the rotating frame 
is described as follows
\begin{align}
    \mathcal{\tilde{H}}(\tau)&= e^{i\tilde{r}\tau\mathcal{H}_{\mathrm{QA}}}\mathcal{H}(\tau)e^{-i\tilde{r}\tau\mathcal{H}_{\mathrm{QA}}}+i\frac{d e^{i\tilde{r}\tau\mathcal{H}_{\mathrm{QA}}}}{d\tau}e^{-i\tilde{r}\tau\mathcal{H}_{\mathrm{QA}}}\nonumber\\
    &=(1-\tilde{r})\mathcal{H}_{\mathrm{QA}} + e^{i\tilde{r}\tau\mathcal{H}_{\mathrm{QA}}}\mathcal{H}_{\mathrm{ext}}(\tau)e^{-i\tilde{r}\tau\mathcal{H}_{\mathrm{QA}}}.\label{transformed_hamil0}
\end{align}
We set
\begin{align}
    \tilde{r}=\frac{\omega}{|E_{m}-E_{0}|}, \label{eq:definition-s}
\end{align} 
where $E_0$ represents the energy eigenvalues
of the ground state. 
The transition frequency between the ground state and the \( m \)-th excited state is assumed to be close to the frequency of the external driving electric field. This means that we have $\tilde{r}\simeq 1$
The second term in Eq.~\eqref{transformed_hamil0} is described as
\begin{align}
&e^{i\tilde{r}\tau\mathcal{H}_{\mathrm{QA}}}\mathcal{H}_{\mathrm{ext}}(\tau)e^{-i\tilde{r}\tau\mathcal{H}_{\mathrm{QA}}}\nonumber\\
   &=\lambda \sum_{i, j}\braket{i|\dot{\mathcal{H}}_{\rm conv}|j}e^{i\tilde{r}(E_{i}-E_{j})\tau}\cos{\omega\tau} \ket{i}\bra{j}\label{eq:inthamil_coeff}
\end{align}
We can simplify this expression as follows. Let us focus on the oscillating part 
in Eq.~\eqref{eq:inthamil_coeff}
such as
\begin{align}
    &\quad e^{i\tilde{r}(E_{i}-E_{j})\tau}\cos{\omega\tau} \nonumber\\ &=\frac{1}{2}e^{i\tilde{r}(E_{i}-E_{j})\tau}(e^{i\omega\tau}+e^{-i\omega\tau}) \nonumber\\
    &=\frac{1}{2}e^{i(\tilde{r}(E_{i}-E_{j})-\omega)\tau}+\frac{1}{2}e^{i(\tilde{r}(E_{i}-E_{j})+\omega)\tau}\label{eq:coeff}
\end{align} 

By using the rotating wave approximation, we obtain

\( H_{\text{ext}, I} = \lambda \langle m| \dot{H}_{\text{conv}}|0 \rangle |m\rangle \langle 0| + \text{h.c.} \) where we use $\tilde{r}\simeq 1$. 
Thus, the effective Hamiltonian of Eq.~\eqref{transformed_hamil0} 
is described as follows:
\begin{align}
    \mathcal{H}_{\mathrm{eff}}= \sum_{i} (1-r)E_{i}\ket{i}\bra{i}+\frac{\lambda}{2}\braket{m|\dot{\mathcal{H}}_{\rm conv}|0}\ket{m}\bra{0} + h.c..\label{eq:external-hamil}
\end{align}
From these calculations, if the initial state is the ground state, the dynamics is limited to the subspace consisting of the ground state and the \( m \)-th excited state. 
Actually, this Hamiltonian is the same as that for the Rabi oscillation. 
Rabi oscillations are generally characterized by detuning and Rabi frequencies. 
Usually, to observe the Rabi oscillations, a projection is made on \( |m\rangle \langle m| \).
The fourth and fifth steps 
are aimed at constructing such a projection measurement. As long as the dynamics of the fourth step is adiabatic, we can construct a projective measurement of \( |m\rangle \langle m| \) in the laboratory system. 
So, we obtain
\begin{align}
p_{0,m}(\omega, s_1, \tau) &\simeq
    |\braket{m|e^{-i\tau\mathcal{H}_{\mathrm{eff}}}|0}|^{2} \nonumber\\
    &\propto  (1-\cos\Omega_{\mathrm{ana}}(\omega)\tau).\label{eq:general-Rabi} 
\end{align}
Here, \( \Omega_{\text{ana}} (\Omega) \) is the Rabi frequency, 
which is defined as 
\begin{align}
    \Omega_{\mathrm{ana}}(\omega) = \sqrt{\left(\lambda|\braket{m|\dot{\mathcal{H}}_{\rm conv}|0}|\right)^{2}+(\omega-\Delta)^{2}},\label{eq:hyperbolic-curve}
\end{align} 
where \( \Delta = E_m - E_0 \) denotes the energy difference between the ground state and $m$-th excited state. 

By applying the analytical expression in Eq.~\eqref{eq:hyperbolic-curve} to the experimental data, 
we can estimate the values of the transition matrix \( | \langle m|\dot{H}|0 \rangle | \) and the energy gap \( E_m - E_0 \)~\cite{mori2022evaluate}.
The 
matrix element \( | \langle m|\dot{H}|0 \rangle |_{\text{est}} \) and the estimated energy gap \( \Delta_{\text{est}} \) are 
given by
\begin{align}
    \lambda |\braket{m|\dot{\mathcal{H}}_{\rm conv}|0}|_{\mathrm{est}}&=\min_{\omega}[\Omega_{\mathrm{exp}}
    (\omega)],\label{eq:numerator_est}\\
        \Delta_{\mathrm{est}}&=
        {\rm{argmin}_{\omega}}
        \left[\Omega_{\mathrm{exp}}(\omega)\right],\label{eq:delta_est}
\end{align}
where $\Omega_{\mathrm{exp}}(\omega)$ is the estimated Rabi frequency obtained from the experiment.
To obtain $\Omega_{\mathrm{exp}}(\omega)$ from the experiment, which can be affected by noise, we can use the following strategy. We calculate the power spectrum as
\begin{align}
    P(\omega,s_{1}, \Omega) &= \mathrm{abs}\left[\mathrm{FT}[p^{({\text{exp}})}_{0,m}(\omega, s_{1}, \tau)]\right],\nonumber\\
    &=\mathrm{abs}\left[\int_{-\infty}^{\infty}d\tau\ p^{({\text{exp}})}_{0,m}(\omega,s_{1},\tau)\frac{e^{-i\Omega\tau}}{\sqrt{2\pi}}\right],\label{def:power-spectrum}
\end{align}
where $p^{({\text{exp}})}_{0,m}(\omega,s_{1},\tau)$ is the data obtained from the experiment. 
If $p^{({\text{exp}})}_{0,m}(\omega,s_{1},\tau)$ coincides with the theoretical form described in \eqref{eq:general-Rabi}, we obtain
\begin{align}
    P(\omega,s_{1},\Omega) \propto &\delta(\Omega) + \frac{1}{2}\delta(\Omega-\Omega_{\mathrm{ana}}(\omega))+\frac{1}{2}\delta(\Omega+\Omega_{\mathrm{ana}}(\omega)).\label{eq:ft-ideal-output}
\end{align}
Since there are some imperfections in the experiments, we do not obtain the delta function in the power spectrum but obtain a broader peak.
So, in the actual experiment, the peak with positive frequency in the spectrum is defined as \( \Omega_{\text{exp}} (\omega) \). We expect that \( \Omega_{\text{exp}} (\omega) \approx \Omega_{\text{ana}} (\omega) \) is satisfied in the power spectrum, which allows us to use Eqs.~\eqref{eq:numerator_est} and \eqref{eq:delta_est}. In this way, the elements of the transition matrix \( | \langle m|\dot{H}_{\text{conv}}|0 \rangle | \) and the energy gap \( \Delta \) can be estimated by the previous method~\cite{mori2022evaluate}.

\section{result}
Here, we explain our method to evaluate the adiabatic condition during QA with the KPOs.
We modify the previous method in Ref.~\cite{mori2022evaluate} as follows.
While the inverse annealing is required in the previous method in the fourth and fifth steps, we replace these steps with a step to measure observable directly.

\begin{figure}[h!]
    \centering
    \includegraphics[width = 8.5cm]{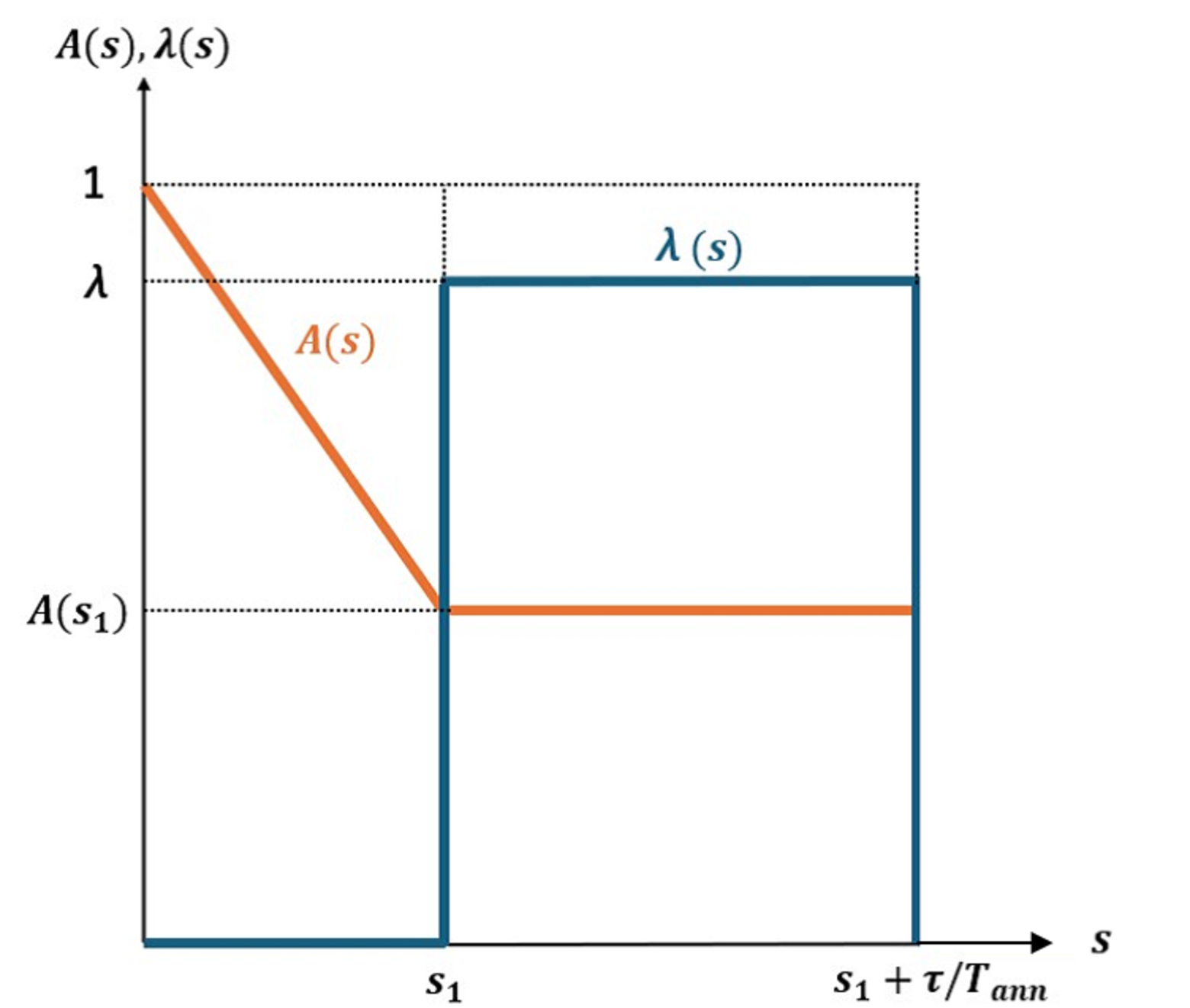}
\caption{
 Scheduling of the 
 parameters in our method to evaluate the adiabatic condition during QA. At $s=s_1+\tau/T_{ann}$, we measure an observable  \( O \).
}\label{zu_posuta.2}
\end{figure}
Specifically, after we perform the step 3, we measure an observable \( \hat{O} \). The candidates for the observable \( \hat{O} \) are \( \hat{x} \), \( \hat{p} \), and \( \hat{n} \) where \( \hat{x} = \hat{a}^\dagger + \hat{a} \), \( \hat{p} = -i(\hat{a}^\dagger - \hat{a}) \), and \( \hat{n} = \hat{a}^\dagger \hat{a} \).
This means that after we perform the third step, we add the following steps.

(4') Fourth, at \( s = s_1 + \frac{\tau}{T_{\text{ann}}} \), \( \lambda(s) = 0 \). 
We turn off the external drive, and measure the observable \( \hat{O} \). 

(5') Finally, these steps are repeated many times by changing \( \omega \), \( s_1 \), and \( \tau \) . We define the expected value of \( \hat{O} \) obtained from these steps as \( \langle \hat{O} \rangle (\omega,s_1,\tau) \).

In Fig.~\ref{zu_posuta.2}, we show the plots of \( A(s) \) and \( \lambda(s) \) with our method. 
The state before we measure the observable  \( \hat{O} \) is approximately given as
\begin{align}
    \cos(\Omega_{\mathrm{ana}}(\omega) \tau)\ket{0(s_{1})}+i\sin(\Omega_{\mathrm{ana}}(\omega) \tau)\ket{m(s_{1})}.
\end{align}
when the frequency of the external field is the same as the energy gap.
So, 
unless the conditions $\langle 0(s_{1})|\hat{O}|0(s_{1})\rangle=\langle m(s_{1})|\hat{O}|m(s_{1})\rangle$ and $\mathrm{Im}[\langle{m(s_{1})|\hat{O}|0(s_{1})}]=0$ are simultaneously satisfied, we can estimate the Rabi frequency from the Fourier transform of the time domain data of \( \langle \hat{O} \rangle (\omega, s_1, \tau) \).
Moreover, after we estimate the Rabi frequency, we can use Eqs.~\eqref{eq:numerator_est} and \eqref{eq:delta_est} for the estimation of the adiabatic condition.

It is worth mentioning that the need for reverse annealing required in the previous method has been eliminated. This is an advantage over the previous method, as reverse annealing takes time to perform adiabatically.
Also, by substituting Eqs.~\eqref{eq:KPO_Hp} and \eqref{eq:KPO_Hd} into Eq.~\eqref{oscillate}, it is shown that we do not need the temporal oscillation of the coupling strength in our method with KPOs. We only need to oscillate the parametric drive amplitude at different frequencies simultaneously, which is feasible, as explained in the Appendix \ref{hiratsukasection}.
 

\subsection{Numerical results for the case of 1 KPO}
\begin{figure}[h!]
    \centering
    \includegraphics[width = 8.5cm]{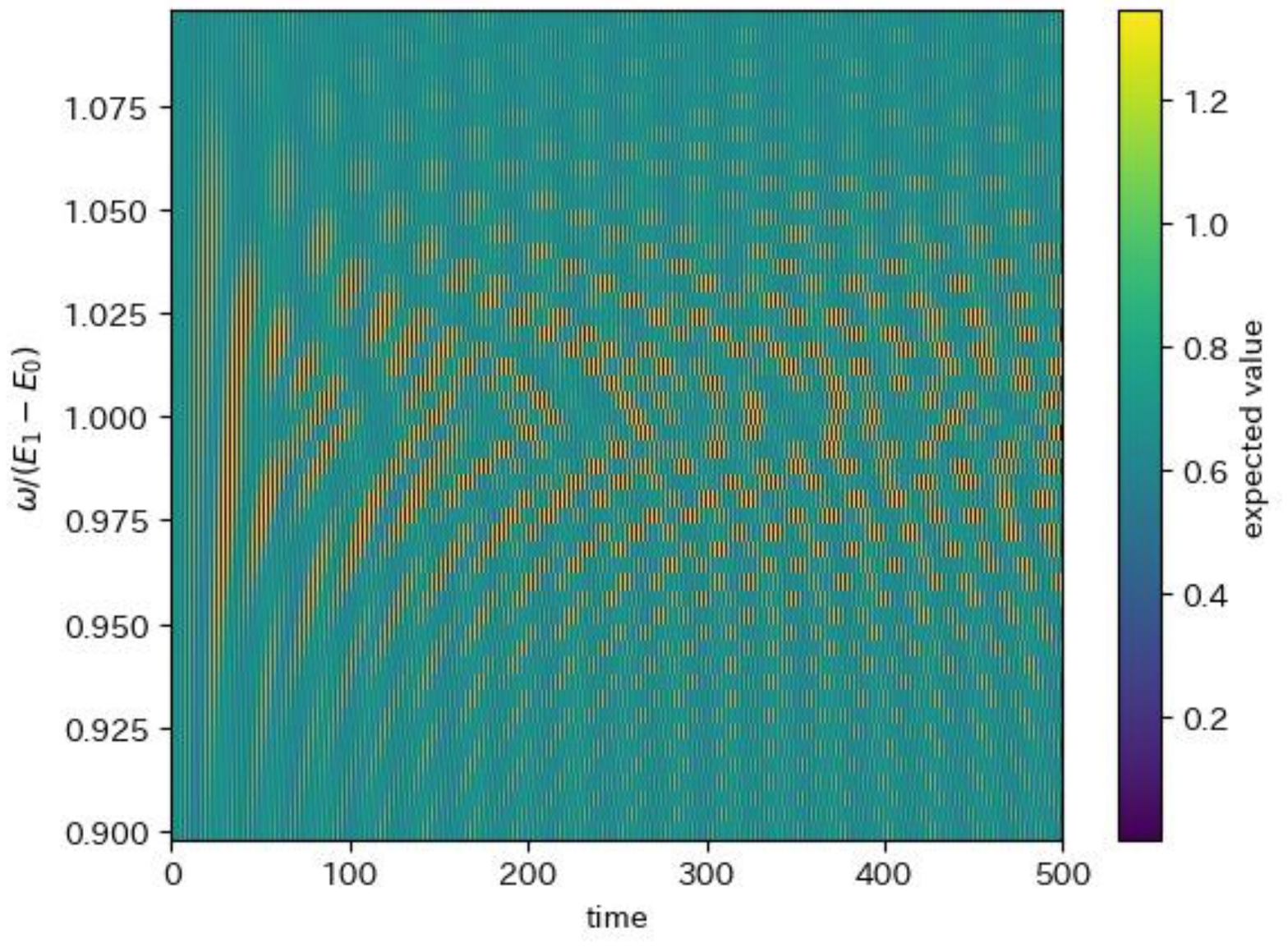}
\caption{
Expected values of $\hat{n}$ when Rabi oscillations are induced by our method. The horizontal axis represents time, the vertical axis shows the 
frequency of the driving field normalized by the energy gap $(E_1 - E_0)$, and the height indicates the expected value. The parameters are set as $K = 1$, $\chi_1 = 1$, $p_1 = 1$, $\Delta_1 = 1$, $r_1 = 1$, $\lambda = 0.1$, $A(s_1) = \frac{1}{2}$, and $T = 500$.
}\label{zu3.1}
\end{figure}

\begin{figure}[h!]
    \centering
    \includegraphics[width = 8.5cm]{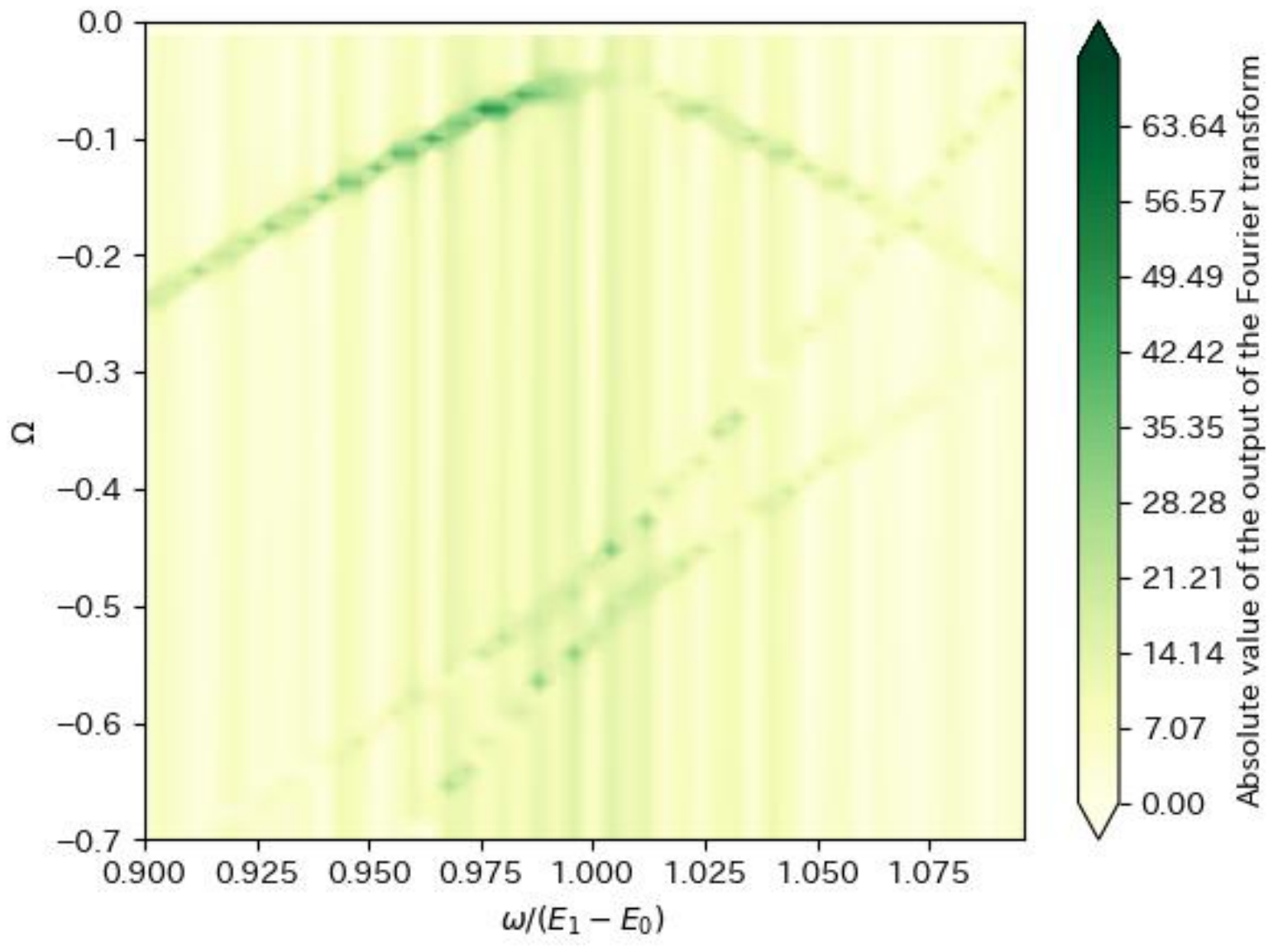}
\caption{
Results of the Fourier transform of Fig. \ref{zu3.1}. 
The horizontal axis is the 
frequency of the driving field normalized by the energy gap $(E_1 - E_0)$, the vertical axis is the frequency of the expected oscillation, and the height is the absolute value of the output after the Fourier transform. 
}\label{zu2.1}
\end{figure}

\begin{figure}[h!]
    \centering
    \includegraphics[width = 8.5cm]{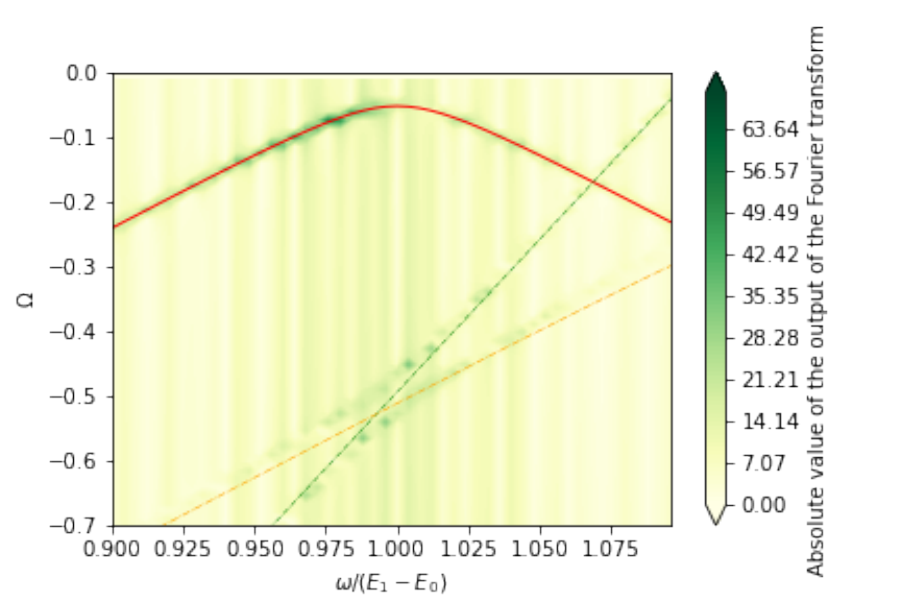}
\caption{
The analytical curves in Eq.~\eqref{eq:hyperbolic-curve}
are added to Fig. \ref{zu2.1}. 
The red continuous line corresponds the target Rabi oscillation between the ground state and first excited state. 
The orange dotted line is analytically expected as that corresponds to the Rabi oscillation between the first and second excited states. The green dotted line corresponds to the oscillation between the ground and second excited states, which arises from the second-order perturbation.} \label{zu1.1}
\end{figure}

First, we numerically evaluate the performance of our method when \( K = 1 \) in Eqs.~\eqref{eq:KPO_Hp} and \eqref{eq:KPO_Hd}.
The parameters are set as \( \chi_1 = 1 \), \( p_1 = 1 \), \( \Delta_1 = 1 \), \( r_1 = 1 \), and \( \lambda = 0.1 \). 
We use \( \hat{n} \) for the observable.
The expected values of the observables at steps 4 and 5 are plotted in Fig.~\ref{zu3.1} when \( \tau \) and \( \omega \) are varied. 
Also, by performing a Fourier transform, we plot
the expected value of the observable 
in the frequency space 
in Fig.~\ref{zu2.1}. 
In Fig.~\ref{zu1.1},
we plot both the numerical results and analytical results with Eq.~\eqref{eq:hyperbolic-curve}. 
Here, for the analytical solution, we diagonalize the Hamiltonian to obtain the transition matrix and energy gap, and we substitute them in Eq.~\eqref{eq:hyperbolic-curve}.

As expected, the numerical results are consistent with the analytical solution.
To evaluate the accuracy of our method, we specifically evaluate the adiabatic condition \eqref{eq:adiabatic_criterion} 
by using Eqs.~\eqref{eq:numerator_est} and \eqref{eq:delta_est}.
The value obtained by diagonalizing the Hamiltonian is $0.096$, whereas the result obtained by our method is $0.091$. 
This shows that our method is effective for the KPO. 

It is notable that two unexpected signals are seen in Figs.~\ref{zu2.1} and \ref{zu1.1}. One of these signals corresponds to another Rabi oscillation between the first and second excited states, which we can fit by using an analytical solution of
$\sqrt{\left(\lambda|\braket{2|\dot{\mathcal{H}}_{\rm conv}|1}|\right)^{2}+(\omega-(E_{2}-E_{0}))^{2}}$ similar to
the Eq.~\eqref{eq:hyperbolic-curve}.
The other signal corresponds to the oscillation between the ground and the second excited state, which arises from the second-order perturbation. We can fit this by using Eq.~\eqref{nondiag} where we substitute $n=0$ and $k=2$ with details provided in Appendix~\ref{app_B}.


\subsection{Numerical results for the case of 2 KPO}

Second, we numerically evaluate the performance of our method when \( K = 2 \) in Eqs.~\eqref{eq:KPO_Hp} and \eqref{eq:KPO_Hd}.
In the time evolution, the following master equation is employed to incorporate the effect of decoherence.
\begin{equation}
    \dot{\rho}(t) = -i[ \mathcal{H}(s),\rho(t)] + \sum_{j=1}^K \frac{1}{2} [2L_j \rho(t) L_j^\dagger - \rho(t) L_j^\dagger L_j - L_j^\dagger L_j \rho(t)]
\end{equation}
where \( L_j = \sqrt{\gamma} \hat{a}_j \) is the decay operator and \( \gamma \) represents the corresponding decoherence rate.
Also, the Hamiltonian of the KPO network commutes with parity operators $e^{i\pi \hat{a}_i^{\dagger}\hat{a}_i}$ for $r_j=0$~\cite{goto2016bifurcation}. 



\begin{figure}[h!]
    \centering
    \includegraphics[width = 8.5cm]{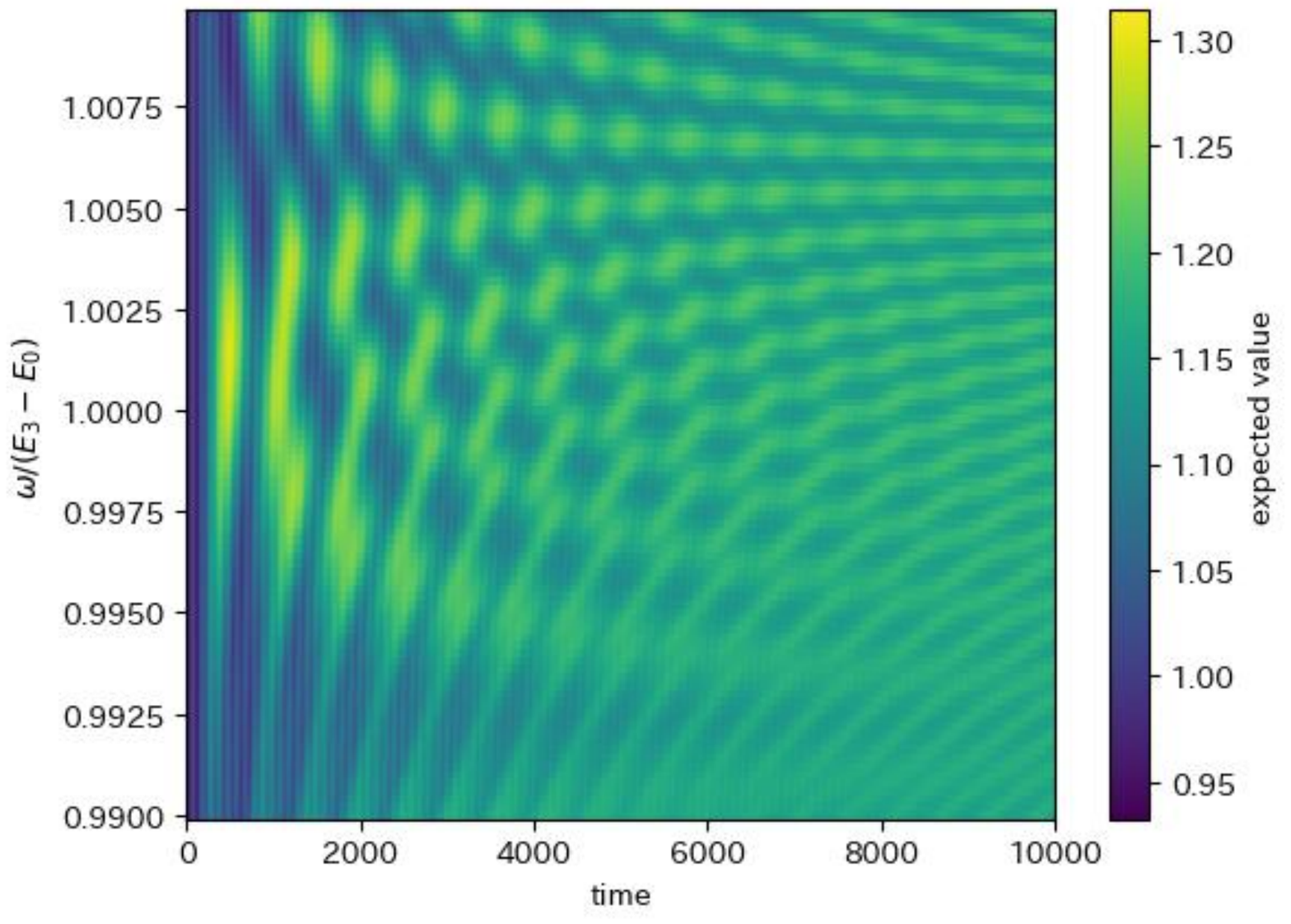}
\caption{Expected value of $\hat{n}_2$ when Rabi oscillations are induced by our method. The horizontal axis represents time, the vertical axis shows the  
frequency of the driving field normalized by the energy gap $(E_3 - E_0)$, and the height indicates the expected value. The parameters used are $T = 500$, $K = 2$, $\chi_1 = 1$, $\chi_2 = 1.23$, $p_1 = 2$, $p_2 = 2.46$, $\Delta_1 = \Delta_2 = 0.1$, $r_1 = r_2 = 0$, $\lambda = 0.02$, $J_{21} = 0.1$, $\gamma = 0.00014$, and $A(s_1) = \frac{1}{3}$.} \label{2KPO_n0}
\end{figure}

In Fig.~\ref{2KPO_n0}, we plot the expected values of the observable at steps 4 and 5 for varying \( \tau \) and \( \omega \) 
For the observable, we use $\hat{n}_2$.

\begin{figure}[h!]
    \centering
    \includegraphics[width = 8.5cm]{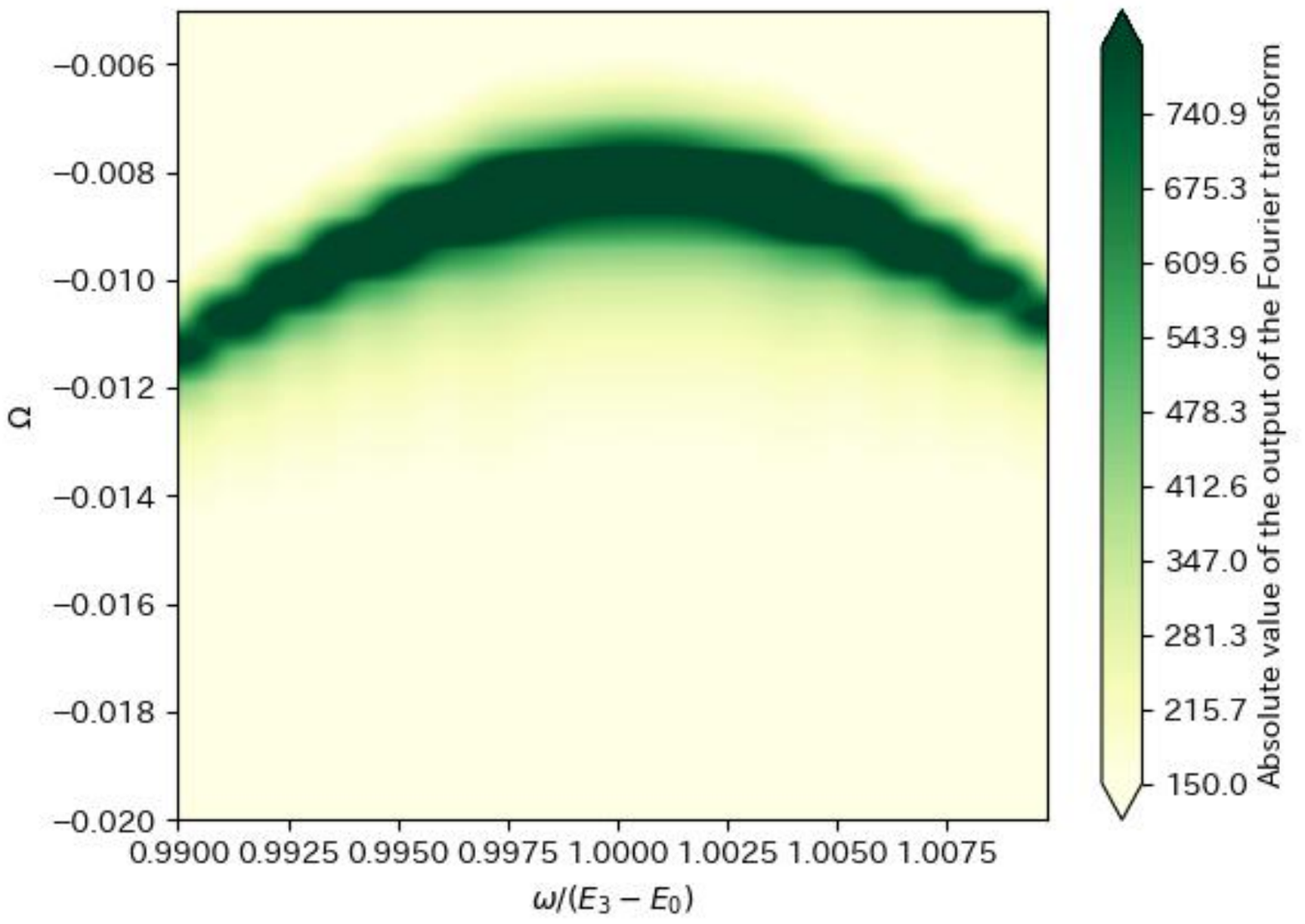}
\caption{Results of the Fourier transform in Fig.~\ref{2KPO_n0}. 
The horizontal axis represents the angular frequency of the added magnetic field normalized by the energy gap $(E_3 - E_0)$, the vertical axis shows the frequency of the expected oscillation, and the height indicates the absolute value of the output after the Fourier transform.
}\label{2KPO_n1}
\end{figure}

Additionally, by performing a Fourier transform, we present 
the expected value of the observable in the frequency domain 
in Fig.~\ref{2KPO_n1}.
In our case, the total Hamiltonian is divided into the odd sector and the even sector of the total number parity $e^{i\pi(a_{1}^{\dagger}a_{1}+a_{2}^{\dagger}a_{2})}$. Since we belong to the even sector in our case, the quantum state does not have any transition to the odd sector. Due to this, we cannot excite the first and second excited states with our method, so we consider a transition between the ground state and the third excited state.

\begin{figure}[h!]
    \centering
    \includegraphics[width = 8.5cm]{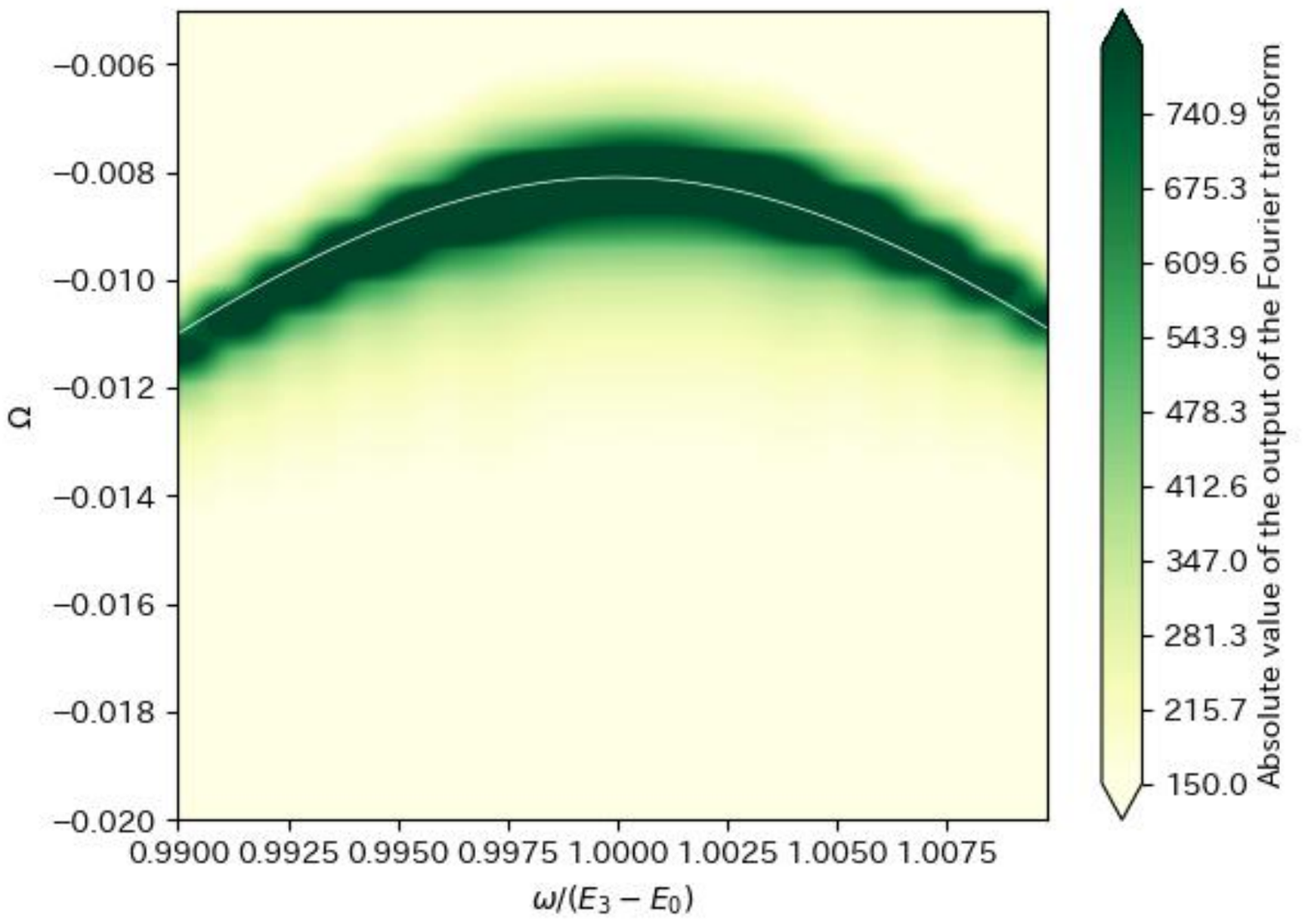}
\caption{An analytical curve in Eq.~\eqref{eq:hyperbolic-curve}
is added to Fig. \ref{2KPO_n1}. 
}\label{2KPO_n2}
\end{figure}

In Fig.~\ref{2KPO_n2}, we present both the numerical and analytical results derived from Eq.~\eqref{eq:hyperbolic-curve}. For the analytical solution, we diagonalize the Hamiltonian to obtain the transition matrix and energy gap, which we then substitute into Eq.~\eqref{eq:hyperbolic-curve}. As anticipated, the numerical results align closely with the analytical solution. To assess the accuracy of our method, we evaluate the adiabatic condition \eqref{eq:adiabatic_criterion} 
by using Eqs.~\eqref{eq:numerator_est} and \eqref{eq:delta_est}. The value obtained through Hamiltonian diagonalization is 0.00816, whereas our method yields a value of  0.00809. So our method is effective for the case of two KPOs with decoherence.

\subsection{2KPO 2}

\section{Conclusion}
We propose a method to estimate the adiabatic condition during QA using Kerr parametric oscillators (KPOs). Unlike previous methods, which require oscillation of the interaction strength between qubits, our approach with KPOs does not necessitate this. This distinction enhances the experimental feasibility of our method. We quantify the performance of our approach and demonstrate its capability to accurately estimate the adiabatic condition, even in the presence of decoherence.


\begin{acknowledgements} 
This paper is partly
based on results obtained from a project, JPNP16007,
commissioned by the New Energy and Industrial Technology Development Organization (NEDO), Japan.

This work was supported by
JST Moonshot (Grant Number JPMJMS226C). Y. Matsuzaki is supported by JSPS KAKENHI (Grant Number
23H04390). This work was also supported by CREST
(JPMJCR23I5), JST.
\end{acknowledgements}

\appendix
\section{Derivation of the Hamiltonian}\label{hiratsukasection} 
Let us consider the case of the single KPO.
The following Hamiltonian is given in the lab frame.
\begin{align}
    \mathcal{H}&=\omega \hat{a}^{\dagger} \hat{a} + \chi (\hat{a}^{\dagger} \hat{a})^2 + p' (\hat{a}^2+\hat{a}^{\dagger 2})\cos{(\omega^{\prime}+\delta)}t\nonumber\\
    &+ p' (\hat{a}^2+\hat{a}^{\dagger 2})\cos{(\omega^{\prime}-\delta)t}
    +2p (\hat{a}^2+\hat{a}^{\dagger 2})\cos(\omega^{\prime}t)
\end{align}
where \( \chi \), \( \Delta \), \( p \) \( (p') \), and \( r \) represent the Kerr nonlinearity, detuning, parametric drive amplitude for the KPO (Rabi oscilation), and coherent drive amplitude, respectively.
Here, we use two parametric drive with different frequencies with $\omega' + \delta$ and $\omega'-\delta$.
We apply the rotational wave approximation to the rotational coordinate system defined by $U=e^{i \frac{\omega^{\prime}}{2}\hat{a}^{\dagger}\hat{a}}$, and we obtain
\begin{align}
    \mathcal{H}_{\rm{eff}}&=UHU^{\dagger}-i\frac{dU}{dt}U^{\dagger}\nonumber\\
    &=(\omega-\omega^{\prime})\hat{a}^{\dagger}\hat{a}+\chi(\hat{a}^{\dagger}a)^2\nonumber\\
    &+p'\frac{e^{i(\omega^{\prime}-\delta)t}+e^{-i(\omega^{\prime}-\delta)t}}{2}(\hat{a}^2e^{-i\omega^{\prime}t}+\hat{a}^{\dagger 2}e^{i\omega^{\prime}t})\nonumber\\
    &+p'\frac{e^{i(\omega^{\prime}+\delta)t}+e^{-i(\omega^{\prime}+\delta)t}}{2}(\hat{a}^2e^{-i\omega^{\prime}t}+\hat{a}^{\dagger 2}e^{i\omega^{\prime}t})\nonumber\\
    &+2p\frac{e^{i\omega^{\prime}t}+e^{-i\omega^{\prime}t}}{2}(\hat{a}^2e^{-i\omega^{\prime}t}+\hat{a}^{\dagger 2}e^{i\omega^{\prime}t})
    \nonumber\\&\approx(\omega-\omega^{\prime})\hat{a}^{\dagger}\hat{a}+\chi(\hat{a}^{\dagger}\hat{a})^2+p'\frac{e^{i\delta t}\hat{a}^2+e^{-i\delta t}\hat{a}^{\dagger 2}}{2}\nonumber\\
    &+p'\frac{e^{-i\delta t}\hat{a}^2+e^{i\delta t}\hat{a}^{\dagger 2}}{2}+2p\frac{\hat{a}^2+\hat{a}^{\dagger 2}}{2}\nonumber\\
    &=(\omega-\omega^{\prime})\hat{a}^{\dagger}\hat{a}+\chi(\hat{a}^{\dagger}\hat{a})^2+p'(\hat{a}^2+\hat{a}^{\dagger 2})\cos{\delta t}\nonumber
    \\
    &+p(\hat{a}^2+\hat{a}^{\dagger 2})
\end{align}
While considering the single KPO, we can generalize this idea to the KPO network and realize the Hamiltonian used in the main text.

\section{Effects of Second-Order Perturbation and the Corresponding Effective Hamiltonian}
\label{app_B}
Up to now, the discussion has been based on effects that can be explained by the rotating wave approximation, that is, by the lowest-order Magnus expansion. However, in practice, effects that cannot be predicted by the rotating wave approximation appear in experiments. Such corrections can be understood as perturbative effects of second order or higher, and the effective Hamiltonian can be constructed from the Magnus expansion of second order or higher. 

First, consider the following time-dependent Hamiltonian,
\begin{align}
    \mathcal{H}(t) = \mathcal{H}_{0} + g\mathcal{H}'\cos{\omega t},
\end{align}
where $\mathcal{H}_{0}$ is an unperturbed Hamiltonian and $g$ is a parameter to denote the strength of the perturbative term. Moving to the interaction picture with a unitary operator $e^{it\alpha\mathcal{H}_{0}}$, this Hamiltonian becomes
\begin{align}
    \mathcal{H}_{R}(t) = ge^{it \alpha\mathcal{H}_{0}}\mathcal{H}'e^{-it\alpha\mathcal{H}_{0}}\cos{\omega t}+(1-\alpha)\mathcal{H}_{0}.\label{Ham_int}
\end{align}
Here, we assume $\alpha\sim 1$ because we consider the case that the detuning is small so that the resonance can be seen.
We insert the identity $I = \sum_{n}|E_{n}\rangle\langle E_{n}|$ where $|E_{n}\rangle$ is the eigenstates of the unperturbed Hamiltonian $\mathcal{H}_{0}$, then the first term of the Hamiltonian~\eqref{Ham_int} reads
\begin{align}
&ge^{it \alpha\mathcal{H}_{0}}\mathcal{H}'e^{-it\alpha\mathcal{H}_{0}}\cos{\omega t}\nonumber\\
    &= g\cos{\omega t}\sum_{n,m}e^{it\alpha\mathcal{H}_{0}}|E_{n}\rangle\langle E_{n}|\mathcal{H}'e^{-it\alpha\mathcal{H}_{0}}|E_{m}\rangle\langle E_{m}|\nonumber\\
    &= g\cos{\omega t}\sum_{n,m} e^{it\alpha(E_{n}-E_{m})}\langle E_{n}|\mathcal{H}'|E_{m}\rangle |E_{n}\rangle\langle E_{m}|\nonumber\\
    &=g\sum_{n,m}\frac{e^{it(\omega+\alpha(E_{n}-E_{m}))}+e^{it(-\omega+\alpha(E_{n}-E_{m}))}}{2}\nonumber \\
    &\qquad \times \langle E_{n}|\mathcal{H}'|E_{m}\rangle |E_{n}\rangle\langle E_{m}|\nonumber \\
    &=g\sum_{n,m} f_{n,m}(t)|E_{n}\rangle\langle E_{m}|,\label{Ham_int_spec}
\end{align}
where 
\begin{align}
    &f_{n,m}(t) \nonumber\\
    &= \frac{e^{it(\omega+\alpha(E_{n}-E_{m}))}+e^{it(-\omega+\alpha(E_{n}-E_{m}))}}{2}\langle E_{n}|\mathcal{H}'|E_{m}\rangle.
\end{align}

We calculate the second order Magnus expansion~\cite{Brinkmann_2016CMR}:
\begin{align}
    \tilde{\mathcal{H}}^{(2)}_{\rm eff}&=\frac{1}{2i(t_{f}-t_{i})}\int^{t_{f}}_{t_{i}} dt_{1}\int_{t_{i}}^{t_{1}} dt[H_{R}(t_{1}),H_{R}(t)]\nonumber\\
    &=\frac{1}{2i(t_{f}-t_{i})}\int^{t_{f}}_{t_{i}} dt_{1}\int_{t_{i}}^{t_{1}} dt\nonumber\\
    &\quad\biggl(\sum_{n,m} (1-\alpha)\left(f_{n,m}(t_1)-f_{n,m}(t)\right)[|E_{n}\rangle\langle E_{m}|,\mathcal{H}_{0}]\nonumber\\
    &\qquad +\sum_{n,m,l,k} f_{n,m}(t_1)f_{l,k}(t)[|E_{n}\rangle\langle E_{m}|,|E_{l}\rangle\langle E_{k}|]\biggr)\nonumber\\
    &=\frac{g(1-\alpha)}{2i(t_{f}-t_{i})}\int^{t_{f}}_{t_{i}} dt_{1}\int_{t_{i}}^{t_{1}} dt \nonumber \\
    &\qquad\times\sum_{n,m} (E_{n}-E_{m})\left(f_{n,m}(t_1)-f_{n,m}(t)\right)|E_{n}\rangle\langle E_{m}|\nonumber\\
    &\quad+\frac{g^{2}}{2i(t_{f}-t_{i})}\int^{t_{f}}_{t_{i}} dt_{1}\int_{t_{i}}^{t_{1}} dt \nonumber \\
    &\qquad\times\sum_{n,m,k} (f_{n,m}(t_1)f_{m,k}(t)-f_{n,m}(t)f_{m,k}(t_1))|E_{n}\rangle\langle E_{k}|.\label{sec_Ham}
\end{align}
Let us evaluate the first term in~\eqref{sec_Ham}. We obtain 
\begin{align}
    &\frac{g}{2i(t_{f}-t_{i})}\int^{t_{f}}_{t_{i}} dt_{1}\int_{t_{i}}^{t_{1}} dt \nonumber \\
    &\qquad\times\sum_{n,m} (E_{n}-E_{m})\left(f_{n,m}(t_1)-f_{n,m}(t)\right)|E_{n}\rangle\langle E_{m}|\nonumber \\
    & =\frac{g}{2i(t_{f}-t_{i})}\int^{t_{f}}_{t_{i}}dt
    \sum_{n,m} (E_{n}-E_{m})\nonumber \\
    &
    ((t_1-t_i)f_{n,m}(t_1)\nonumber \\
    &-\frac{e^{it_1(\omega+\alpha(E_{n}-E_{m}))}-e^{it_i(\omega+\alpha(E_{n}-E_{m}))}}{2i(\omega+\alpha(E_{n}-E_{m}))}\langle E_{n}|\mathcal{H}'|E_{m}\rangle \nonumber \\
    &- \frac{e^{it_1(-\omega+\alpha(E_{n}-E_{m}))}-e^{it_i(-\omega+\alpha(E_{n}-E_{m}))}}{2i(-\omega+\alpha(E_{n}-E_{m}))}\langle E_{n}|\mathcal{H}'|E_{m}\rangle
     )\nonumber \\
     &
     |E_{n}\rangle\langle E_{m}|\nonumber \\
       & =\frac{g}{2i(t_{f}-t_{i})}
    \sum_{n,m} (E_{n}-E_{m})\nonumber \\
    &\Big{(}
    \frac{e^{it(\omega+\alpha (E_n-E_m))}}{2}(\frac{1}{(\omega+\alpha (E_n-E_m))^2}-\frac{it}{\omega+\alpha (E_n-E_m)})\nonumber \\
    &+
    \frac{e^{it(\omega-\alpha (E_n-E_m))}}{2}(\frac{1}{(\omega-\alpha (E_n-E_m))^2}-\frac{it}{(\omega-\alpha (E_n-E_m))})+
\nonumber \\
    &
    \int^{t_{f}}_{t_{i}}dt(-t_if_{n,m}(t_1)\nonumber \\
    &-\frac{e^{it_1(\omega+\alpha(E_{n}-E_{m}))}-e^{it_i(\omega+\alpha(E_{n}-E_{m}))}}{2i(\omega+\alpha(E_{n}-E_{m}))}\langle E_{n}|\mathcal{H}'|E_{m}\rangle \nonumber \\
    &- \frac{e^{it_1(-\omega+\alpha(E_{n}-E_{m}))}-e^{it_i(-\omega+\alpha(E_{n}-E_{m}))}}{2i(-\omega+\alpha(E_{n}-E_{m}))}\langle E_{n}|\mathcal{H}'|E_{m}\rangle
     )\nonumber \\
     &
     |E_{n}\rangle\langle E_{m}|\Big{)}
\end{align}
So, by assuming $g \gg |\pm\omega +\alpha(E_n - E_m)|$, we can ignore this term where we use Eq.~\eqref{moriformula}.


To evaluate Eq.~\eqref{sec_Ham}, we calculate
\begin{align}
    &\frac{1}{t_{f}-t_{i}}\int^{t_{f}}_{t_{i}} dt_{1}\int_{t_{i}}^{t_{1}} dt (f_{n,m}(t_1)f_{m,k}(t)-f_{n,m}(t)f_{m,k}(t_1))\nonumber\\
    &=\frac{\langle E_{n}|\mathcal{H}'|E_{m}\rangle\langle E_{m}|\mathcal{H}'|E_{k}\rangle}{4(t_{f}-t_{i})}\int^{t_{f}}_{t_{i}} dt_{1}\nonumber\\
    &\biggl(\frac{e^{it_{1}(2\omega+\alpha(E_{n}-E_{k}))}+e^{it_{1}\alpha(E_{n}-E_{k})}}{i(\omega+\alpha(E_{m}-E_{k}))}\nonumber \\
    &-\frac{e^{it_{i}(\omega+\alpha(E_{m}-E_{k}))}(e^{it_{1}(\omega+\alpha(E_{n}-E_{m}))}+e^{it_{1}(-\omega+\alpha(E_{n}-E_{m}))})}{i(\omega+\alpha(E_{m}-E_{k}))}\nonumber\\
    &+\frac{e^{it_{1}(-2\omega+\alpha(E_{n}-E_{k}))}+e^{it_{1}\alpha(E_{n}-E_{k})}}{i(-\omega+\alpha(E_{m}-E_{k}))}\nonumber \\
    &-\frac{e^{it_{i}(-\omega+\alpha(E_{m}-E_{k}))}(e^{it_{1}(\omega+\alpha(E_{n}-E_{m}))}+e^{it_{1}(-\omega+\alpha(E_{n}-E_{m}))})}{i(-\omega+\alpha(E_{m}-E_{k}))}\nonumber\\
    &-\frac{e^{it_{1}(2\omega+\alpha(E_{n}-E_{k}))}+e^{it_{1}\alpha(E_{n}-E_{k})}}{i(\omega+\alpha(E_{n}-E_{m}))}\nonumber \\
    &+\frac{e^{it_{i}(\omega+\alpha(E_{n}-E_{m}))}(e^{it_{1}(\omega+\alpha(E_{m}-E_{k}))}+e^{it_{1}(-\omega+\alpha(E_{m}-E_{k}))})}{i(\omega+\alpha(E_{n}-E_{m}))}\nonumber\\
    &-\frac{e^{it_{1}(-2\omega+\alpha(E_{n}-E_{k}))}+e^{it_{1}\alpha(E_{n}-E_{k})}}{i(-\omega+\alpha(E_{n}-E_{m}))}\nonumber \\
    &+\frac{e^{it_{i}(-\omega+\alpha(E_{n}-E_{m}))}(e^{it_{1}(\omega+\alpha(E_{m}-E_{k}))}+e^{it_{1}(-\omega+\alpha(E_{m}-E_{k})}))}{i(-\omega+\alpha(E_{n}-E_{m}))}\biggr).\label{integr_2}
\end{align}
Here, we apply the large $t_{f}-t_{i}$ limit. All the terms in Eq.~\eqref{integr_2} are proportional to,
\begin{align}
\frac{1}{t_{f}-t_{i}}\int_{t_{i}}^{t_{f}} dt_{1} e^{ict_{1}}=\quad\ &\begin{cases}
    \frac{e^{ict_{f}}-e^{ict_{i}}}{ic(t_{f}-t_{i})}& (c\neq 0)\\
    1 & (c= 0)
\end{cases},\nonumber\\
\overset{t_{f}-t_{i}\to\infty}{=}
&\begin{cases}
    0 & (c\neq 0)\\
    1 & (c= 0).\label{moriformula}
\end{cases}
\end{align}
This means that we can neglect terms with a condition $c\neq 0$. 

We can find nonzero off-diagonal ($n \neq k$) terms when the condition, 
\begin{align}
    2\omega + \alpha(E_{n}-E_{k}) = 0,\label{cond_al}
\end{align}
is satisfied. We choose a pair of $(n,k)$ 
to satisfy the condition~\eqref{cond_al}.
The nonzero off-diagonal terms are given by
\begin{align}
     &\frac{\langle E_{n}|\mathcal{H}'|E_{m}\rangle\langle E_{m}|\mathcal{H}'|E_{k}\rangle}{4(t_{f}-t_{i})}\int_{t_{i}}^{t_{f}} dt_{1}\nonumber\\
     & \left(\frac{1}{i(\omega+\alpha(E_{m}-E_{k}))}-\frac{1}{i(\omega+\alpha(E_{n}-E_{m}))}\right),\nonumber\\
    &=\frac{\langle E_{n}|\mathcal{H}'|E_{m}\rangle\langle E_{m}|\mathcal{H}'|E_{k}\rangle(E_{k}-E_{n})}{2i\omega(2E_{m}-E_{n}-E_{k}))}.
    \label{amp_oscil} 
\end{align}


Finally, the effective Hamiltonian becomes
\begin{align}
    \mathcal{H}_{\rm eff} = \left(1-\frac{2\omega}{E_{k}-E_{n}}\right)\mathcal{H}_{0}
    +\mathcal{H}^{\rm nd}_{\rm eff}.\label{semfin2ham}
\end{align}


To study the frequency of the oscillation between the state $|E_{n}\rangle$ and $|E_{k}\rangle$, we project out this effective Hamiltonian into the subspace spanned by $|E_{n}\rangle$ and $|E_{k}\rangle$. Then, the Hamiltonian becomes
\begin{align}
    & \langle E_{n}|\mathcal{H}_{\rm eff}^{\rm nd}|E_{k}\rangle|E_{n}\rangle\langle E_{k}|+h.c.\nonumber\\
    &+ \left(\left(1-\frac{2\omega}{E_k-E_n}\right)E_{n}\right)|E_{n}\rangle\langle E_{n}|\nonumber\\
    &+\left(\left(1-\frac{2\omega}{E_k-E_n}\right)E_{k}\right)|E_{k}\rangle\langle E_{k}|,
\end{align}
where 
\begin{align}
    \langle E_{n}|\mathcal{H}_{\rm eff}^{\rm nd}|E_{k}\rangle =-\frac{g^{2}}{2}\sum_{m}\frac{\langle E_{n}|\mathcal{H}'|E_{m}\rangle\langle E_{m}|\mathcal{H}'|E_{k}\rangle}{\frac{2\omega}{E_{k}-E_{n}}(2E_{m}-E_{n}-E_{k}))}
\end{align}

Compared to the case of the Rabi oscillation in the qubit system, the angular frequency of the Rabi oscillation is given by,
\begin{align}
    \Omega_{\rm ana}(\omega)=\sqrt{(E_{k}-E_{n}-2\omega)^{2}+|\langle E_{n}|\mathcal{H}_{\rm eff}^{\rm nd}|E_{k}\rangle|^{2}}.\label{nondiag}
\end{align}

\bibliography{bib_main}

\end{document}